\documentclass{aa} 
\usepackage{txfonts}
\usepackage{longtable}  
\usepackage{rotating}
\usepackage{natbib}
\usepackage{graphicx}
\usepackage{graphics}
\usepackage{psfrag}
\usepackage{amssymb}
\usepackage{multicol}
\usepackage{lscape}
\bibpunct{(}{)}{;}{a}{}{,}

\def\vsini{\ensuremath{{\upsilon}\sin i}}

\def\kms{$\mathrm{km\,s}^{-1}$}

\def\nlte{non-LTE}

\def\bz{$\langle$B$_z\rangle$}
\def\nz{$\langle$N$_z\rangle$}
\def\sbz{$\sigma_{\langle\,B_z\,\rangle}$}
\def\snz{$\sigma_{\langle\,N_z\,\rangle}$}
\begin{document} 
\title{B fields in OB stars (BOB): low-resolution FORS2 spectropolarimetry of the first sample of 50 massive stars\thanks{Based on observations made with ESO Telescopes at the La Silla Paranal Observatory under programme ID\,191.D-0255(A,C).}} 
\author{L. Fossati\inst{1,2}	\and
	N. Castro\inst{2}	\and
	M. Sch\"oller\inst{3}   \and
	S. Hubrig\inst{4}       \and
	N. Langer\inst{2}	\and
	T. Morel\inst{5}      \and
	M. Briquet\inst{5}    \fnmsep\thanks{F.R.S.-FNRS Postdoctoral 								Researcher, Belgium}\and
	A. Herrero\inst{6,7}   \and
	N.~Przybilla\inst{8}   \and
	H. Sana\inst{9}		   \and
	F.~R.~N. Schneider\inst{10,2}	\and
	A. de Koter\inst{11,12}	\and
	the BOB collaboration
}
\institute{
	Space Research Institute, Austrian Academy of Sciences, Schmiedlstrasse 		6, A-8042 Graz, Austria\\
	\email{luca.fossati@oeaw.ac.at}
	\and 
	Argelander-Institut f\"ur Astronomie der Universit\"at Bonn, Auf dem 			H\"ugel 71, 53121, Bonn, Germany
	\and 
	European Southern Observatory, Karl-Schwarzschild-Str. 2, D-85748 			Garching, Germany
	\and
	Leibniz-Institut f\"ur Astrophysik Potsdam (AIP), An der Sternwarte 16, 		D-14482 Potsdam, Germany
	\and
	Institut d'Astrophysique et de G\'eophysique, Universit\'e de Li\`ege, 			All\'ee du 6 Ao\^ut, B\^at. B5c, 4000 Li\`ege, Belgium
	\and
	Instituto de Astrof\'isica de Canarias, C/ V\'ia L\'actea s/n, 38200 La 		Laguna, Tenerife, Spain
	\and
	Departamento de Astrof\'isica, Universidad de La Laguna, Avda. 				Astrof\'isico Francisco S\'anchez s/n, 38071 La Laguna, Tenerife, Spain
	\and
	Institut f\"ur Astro- und Teilchenphysik, Universit\"at Innsbruck, 			Technikerstr. 25/8, 6020 Innsbruck, Austria
	\and
	European Space Agency, Space Telescope Science Institute, 3700 San 			Martin Drive, Baltimore, MD 21218, USA
	\and
	Department of Physics, Denys Wilkinson Building, Keble Road, Oxford, OX1 	3RH, United Kingdom
	\and
	Astronomical Institute Anton Pannekoek, University of Amsterdam, 			Kruislaan 403, 1098 SJ, Amsterdam, The Netherlands
	\and
	Instituut voor Sterrenkunde, KU Leuven, Celestijnenlaan 200D, 3001, 			Leuven, Belgium
} 
\date{} 
\abstract
{
Within the context of the collaboration ``B fields in OB stars (BOB)'', we used the FORS2 low-resolution spectropolarimeter to search for a magnetic field in 50 massive stars, including two reference magnetic massive stars. Because of the many controversies of magnetic field detections obtained with the FORS instruments, we derived the magnetic field values with two completely independent reduction and analysis pipelines. We compare and discuss the results obtained from the two pipelines. We obtaind a general good agreement, indicating that most of the discrepancies on magnetic field detections reported in the literature are caused by the interpretation of the significance of the results (i.e., 3--4$\sigma$ detections considered as genuine, or not), instead of by significant differences in the derived magnetic field values. By combining our results with past FORS1 measurements of HD\,46328, we improve the estimate of the stellar rotation period, obtaining P\,=\,2.17950$\pm$0.00009\,days. For HD\,125823, our FORS2 measurements do not fit the available magnetic field model, based on magnetic field values obtained 30 years ago. We repeatedly detect a magnetic field for the O9.7V star HD\,54879, the HD\,164492C massive binary, and the He-rich star CPD\,$-$57\,3509. We obtain a magnetic field detection rate of 6$\pm$4\%, while by considering only the apparently slow rotators we derive a detection rate of 8$\pm$5\%, both comparable with what was previously reported by other similar surveys. We are left with the intriguing result that, although the large majority of magnetic massive stars is rotating slowly, our detection rate is not a strong function of the stellar rotational velocity.
}
\keywords{Stars: magnetic field -- Stars: early-type -- Stars: massive}
\titlerunning{B fields in OB stars (BOB): low resolution FORS2 spectropolarimetry of 50 massive stars}
\authorrunning{L. Fossati et al.}
\maketitle
\section{Introduction}\label{sec:introduction}
Magnetic fields play an important role in the structure and evolution of stars, and systematic surveys aiming at the detection and characterisation of magnetic fields in massive stars, have only recently started to be carried out \citep{wade2014,morel2014,morel2015}. Their most evident achievement is the great increase in the number of detected magnetic massive stars, leading for example to the determination of a magnetic field incidence of $\sim$7\%, made on the basis of a sample of hundreds of stars \citep{wade2014}. Recently, the detection of rather weak magnetic fields opened the possibility that the incidence may be higher, calling for deeper observations for the brightest stars \citep{fossati2015}. 

Despite these achievements, the number of known magnetic massive stars is still relatively small, particularly with respect to the wide variety of detected phenomena and features in their spectra and light curves. The detection of more magnetic massive stars is therefore a necessary step for further advances.

This work is part of the collaboration ``B fields in OB stars'' (BOB), whose primary aim is to characterise the incidence of large-scale magnetic fields in slowly rotating (i.e., \vsini\,$\lesssim$\,100\,\kms) main-sequence massive stars (i.e., early B- and O-type stars), to test whether the slow rotation is primarily caused by the presence of a magnetic field. The observations are being performed with the high-resolution HARPSpol polarimeter \citep{snik2011,piskunov2011}, feeding the HARPS spectrograph \citep{mayor2003} attached to the ESO 3.6\,m telescope in La\,Silla (Chile), and the FORS2 low-resolution spectropolarimeter \citep{app1992} attached to the Cassegrain focus of the 8\,m Antu telescope of the ESO Very Large Telescope of the Paranal Observatory. More details about the BOB collaboration can be found in \citet{morel2014,morel2015}. We present here the results obtained from the first set of 50 stars, while the results of a subsequent sample will be presented in a forthcoming paper (Sch\"oller et al., in preparation).
\section{Target selection}\label{sec:targ_selection}
The target selection was performed considering the stellar {\it i})~spectral type (O- and early B-type stars), {\it ii}) luminosity class (dwarfs and giants; V$\rightarrow$III), and {\it iii}) projected rotational velocity (\vsini\,$\leq$\,100\,\kms). As main sources of information we used \citet{howarth1997}, the UVES Paranal Observatory Project spectral library \citep[because of the availability of high-resolution spectra, which would in particular complement the low-resolution FORS2 observations;][]{uvespop}, the GOSSS survey \citep[][Barba priv. communication]{gosss}, and the IACOB database \citep{iacob}. We also checked the catalogue compiled by \citet{bychkov2009} for previous magnetic field measurements, while we gathered information about possible binarity from the surveys cited above. As shown by \citet{babel1997}, the interaction of the stellar wind of magnetic massive stars with their magnetosphere can be a strong source of hard X-rays, which may be detectable if the stars are close enough. For this reason, we included in our target list previously identified hard X-ray sources, using available X-ray catalogues and archival X-ray data, with \vsini\ values up to 120\,\kms.

The selected sample of stars also includes two known magnetic reference stars: HD\,46328 \citep{hubrig2006} and HD\,125823 \citep{wolff1974,borra1983}. We tried to limit the observations of supergiants (luminosity class I) because for these we cannot exclude that even a non-magnetic wind \citep{langer1998} might have spun them down. The compiled target list was then split according to stellar magnitude, so that stars with $V\gtrsim$\,7.5\,mag have been preferentially observed with FORS2 and the remaining with HARPSpol.
\section{Observations}\label{sec:observations}
FORS2 is a multi-mode optical instrument capable of imaging, polarimetry, and long-slit and multi-object spectroscopy. The polarimetric optics, previously mounted on FORS1 \citep{app1998}, have been moved to FORS2 in March 2009. During the first run, performed between the 7 and 9 of April 2013, we observed 24 stars, while during the second run, performed between the 6 and 8 of February 2014, we observed 28 stars (HD\,102475 and HD\,144470 were observed during both runs). The observing log of both runs is given in Table~\ref{tab:obs.log}.

For the first run, we used the 2k$\times$4k E2V CCDs (pixel size 15\,$\mu$m\,$\times$\,15\,$\mu$m) which are optimised for observations in the blue spectral region (i.e., $<$4500\,\AA), while for the second run we used the 2k$\times$4k MIT CCDs (pixel size 15\,$\mu$m\,$\times$\,15\,$\mu$m)\footnote{The E2V CCDs have a nominal gain (conversion from counts to electrons) of 2.20 and a readout noise (in electrons) of 4.20, while the MIT CCDs have a nominal gain of 1.25 and a readout noise of 2.70.}. All observations were performed using a single narrow slit width of 0.4$\arcsec$, to reach a high spectral resolution and to minimise spurious effects of seeing variations \citep[see e.g.][]{fossati2015b}, the 200\,kHz/low/1$\times$1 readout mode, to minimise overheads and increase the dynamic range, and the GRISM\,600B. Each spectrum covers the 3250--6215\,\AA\ spectral range which includes all Balmer lines, except H$\alpha$, and a number of He lines. Using the emission lines of the wavelength calibration lamp we measured an average (across the covered wavelength range) resolving power of 1700. Each star was observed with a sequence of spectra obtained by rotating the quarter waveplate alternatively from $-$45$^{\circ}$ to $+$45$^{\circ}$ every second exposure (i.e., $-$45$^{\circ}$, $+$45$^{\circ}$, $+$45$^{\circ}$, $-$45$^{\circ}$, $-$45$^{\circ}$, $+$45$^{\circ}$, etc.). The adopted exposure times and obtained signal-to-noise ratios (S/N) per pixel calculated around 4950\,\AA\ of Stokes $I$ are listed in Table~\ref{tab:obs.log}.
\section{Data reduction and analysis}\label{reduction.analysis}
Because of the several controversies present in the literature about magnetic field detections in intermediate- and high-mass stars performed with the FORS spectropolarimeters \citep[see e.g. ][]{wade2007,silvester2009,shultz2012,bagnulo2012,bagnulo2013}, the data were independently reduced by two different groups (one based in Bonn and one based in Potsdam) using a set of completely independent tools and routines. The first reduction and analysis (Bonn) was performed with a set of IRAF\footnote{Image Reduction and Analysis Facility (IRAF -- {\tt http://iraf.noao.edu/}) is distributed by the National Optical Astronomy Observatory, which is operated by the Association of Universities for Research in Astronomy (AURA) under cooperative agreement with the National Science Foundation.} \citep{tody} and IDL routines (hereafter called Bonn pipeline) developed following most of the technique and recipes presented by \citet{bagnulo2012,bagnulo2013}, while the second reduction and analysis (hereafter called Potsdam pipeline) was based on the tools described in \citet{hubrig2004a,hubrig2004b}, with the recent update described in \citet{steffen2014}.

The surface-averaged longitudinal magnetic field \bz\ was measured using the following relation \citep{angel1970,landstreet1975}:
\begin{equation}
\label{eq:bz}
V(\lambda)=-g_{\rm eff}C_z\lambda^2\frac{1}{I(\lambda)}\frac{{\rm d}I(\lambda)}{{\rm d}\lambda}\langle\,B_z\,\rangle
\end{equation}
and the least-squares technique, originally proposed by \citet{bagnulo2002} and further refined by \citet{bagnulo2012}. In Eq.~\ref{eq:bz} $V(\lambda)$ and $I(\lambda)$ are the Stokes $V$ and $I$ profiles, respectively, $g_{\rm eff}$ is the effective Land\'e factor, which was set to 1.25 except for the region of the hydrogen Balmer lines where $g_{\rm eff}$ was set to 1.0, and 
\begin{equation}
\label{eq:cz}
C_z=\frac{e}{4 \pi m_ec^2}
\end{equation}
where $e$ is the electron charge, $m_e$ the electron mass, and $c$ the speed of light ($C_z\simeq4.67\,\times\,10^{-13}$\,\AA$^{-1}$G$^{-1}$). See \citet{bagnulo2012} for a detailed discussion of the physical limitations of this technique.

In the remainder of this section, we thoroughly describe the routines and settings adopted within the two pipelines. We also schematically summarise the main similarities and differences.
\begin{table*}[]
\caption[ ]{Comparison between the reduction and analysis procedures applied within the two adopted pipelines. More details are given in the text.}
\label{tab:pipelines}
\begin{center}
\begin{tabular}{l|c|c}
\hline
\hline
Reduction/analysis step & Bonn pipeline & Potsdam pipeline \\
\hline
Bias subtraction                 & Yes             & Yes         \\
Flat-field correction            & No              & No          \\
Spectral extraction method       & Average         & Average     \\
Background subtraction           & No              & Yes         \\
Wavelength calibration           & Manual          & Automatic   \\
Spectral sampling [\AA/pix]      & 0.75            & 0.1         \\
Rectification                    & Polynomial      & Linear      \\
Sigma clipping                   & Yes             & Yes         \\
\bz, \nz                         & Linear fit      & Linear fit  \\
\sbz, \snz                       & $\chi^2$-scaled & Monte Carlo \\
\hline
\end{tabular}
\end{center}
\end{table*}

%
\subsection{Bonn pipeline}
Within the Bonn pipeline, we applied a bias subtraction, but no flat-field correction\footnote{For polarisation measurements, from a mathematical point of view the flat-field correction has no influence on the results. However, \citet{bagnulo2012} showed that in practice this is not the case, most likely because of fringing, but it is not possible to clearly identify the best option.}. We performed an average extraction, as recommended by \citet{bagnulo2012}, using a fixed extraction radius of 25 pixels, without background subtraction. The adopted extraction radius allowed us to avoid the spectrum of the parallel beam being contaminated by a strong instrumental internal reflection, which would otherwise irreparably affect the Stokes profiles in the region around H$\delta$. Within each night, each parallel or perpendicular beam was wavelength calibrated using the parallel or perpendicular beam of one wavelength calibration lamp obtained in the morning following the night of observation. The wavelength calibration was performed manually to ensure that the same set of arc lines and fitting functions were used for both beams \citep{bagnulo2013}. The pipeline finally bins the spectra according to the natural sampling of the instrument/grism of 0.75\,\AA/pix.

We combined the profiles to obtain Stokes $I$, $V$, and the diagnostic $N$ parameter \citep{donati1992} using the difference method following the formalism of \citet{bagnulo2009}\footnote{Optionally, the IDL routine allows one to calculate the uncertainty of Stokes $V$ using the simplified formulation given in Eq.~A6 of \citet{bagnulo2009}, which is valid for low polarisation values.}. We rectified each Stokes $V$ profile using a fourth-order polynomial and applied a sigma clipping to filter out all data points where the $N$ profile deviated more than 3$\sigma$ from the average value ($\overline{N}$), where $\sigma$ is the standard deviation of the $N$ profile. The value of \bz\ was calculated using either the hydrogen lines, the metallic lines, or the whole spectrum in the 3710--5870\,\AA\ spectral region. The Stokes $I$ spectra were inspected to remove all spectral regions contaminated by emission lines. The field was calculated minimising
\begin{equation}
\label{eq:chi}
\chi^2=\sum_i\frac{(V(\lambda_i)-\langle\,B_z\,\rangle\,x_i-b)^2}{\sigma_i^2}
\end{equation}
where $x_i$\,=\,$-g_{\rm eff}C_z\lambda_i^2(1/I(\lambda)\times{\rm d}I(\lambda)/{\rm d}\lambda)_i$, $i$ indicates each spectral point, and $b$ is a constant that accounts for possible spurious continuum polarisation left after the rectification \citep[see][for more details]{bagnulo2002,bagnulo2012}. Finally, the code provides the values of \bz\ and \nz\  (the magnetic field calculated from the $N$ profile), their standard uncertainty, and their $\chi^2$-scaled uncertainty \citep[\sbz\ and \snz\ -- see Sect.~3.4 of][]{bagnulo2012}. Optionally, the IDL routine allows one to extract \bz\ with a $\chi^2$ minimisation routine that takes into account the uncertainties on both axes, using the {\sc astrolib fitexy.pro}\footnote{\tt http://idlastro.gsfc.nasa.gov/} routine based on a routine that is part of the numerical recipes \citep{press1992}. In this work, we always adopted the $\chi^2$-scaled uncertainties, taking into account only the error bars on Stokes $V$. By adopting the $\chi^2$-scaled uncertainties, we also compensated for variations of the CCD gain from the nominal value, which was adopted for the spectral extraction. Using the bias and flat-field calibration frames collected during our runs, we consistently measured a CCD gain slightly lower than the adopted nominal value. This is confirmed, for example, by the fact that for the $N$ profile we constantly obtained an average uncertainty smaller than the standard deviation. 
\subsection{Potsdam pipeline}
%
\begin{figure*}[ht!]
\begin{center}
\includegraphics[width=180mm]{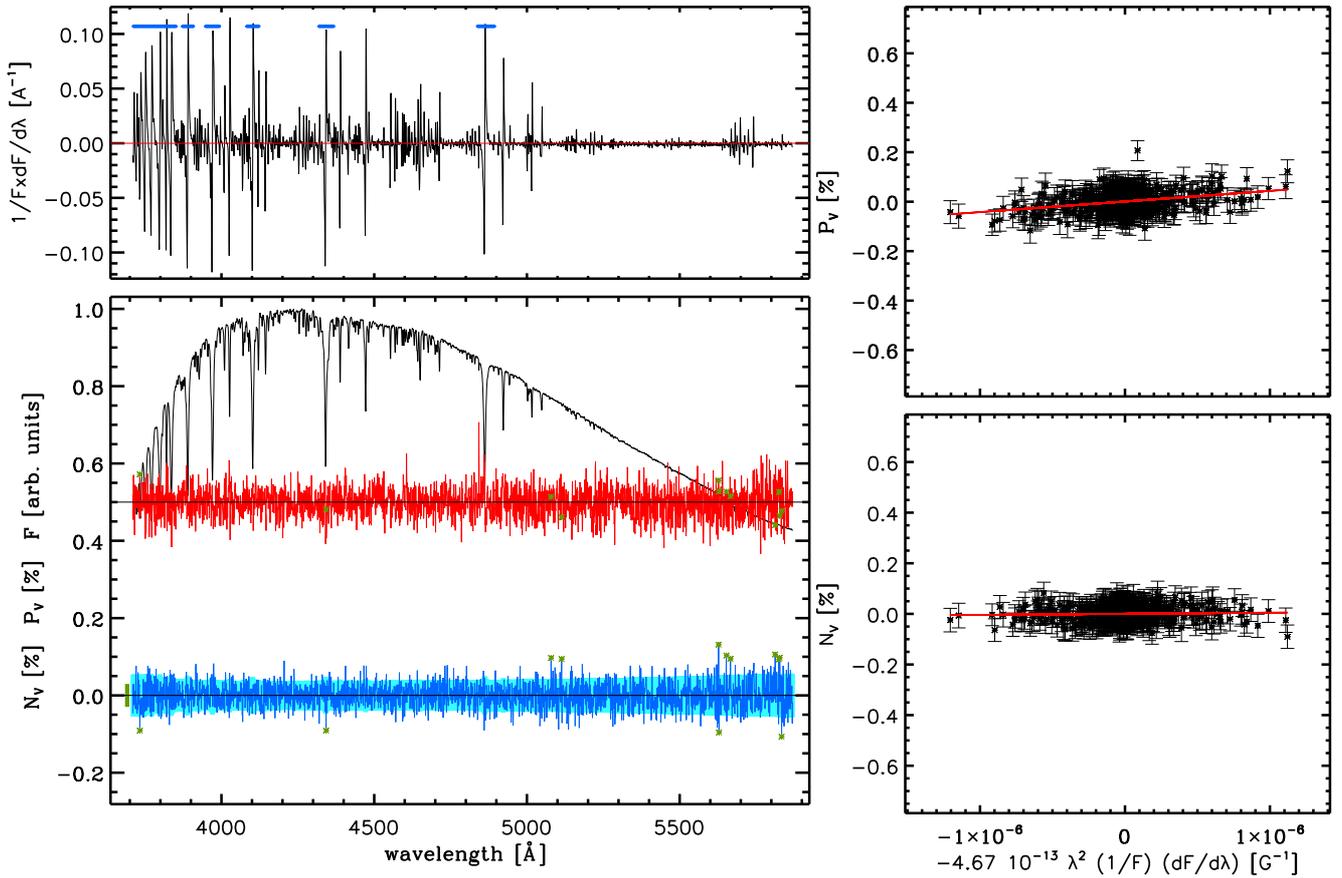}
\caption{Overview of the results of the analysis of the FORS2 data of HD\,46328, collected on 7 April 2013, considering the hydrogen lines, using the Bonn pipeline. Top left panel: derivative of Stokes $I$. The regions used to calculate the magnetic field are marked by a thick blue line close to the top of the panel. Bottom left panel: the top profile shows Stokes $I$ arbitrarily normalised to the highest value, the middle red profile shows Stokes $V$ (in \%) rigidly shifted upwards by 0.5\% for visualisation reasons, while the bottom blue profile shows the spectrum of the $N$ parameter (in \%). The green asterisks mark the points that were removed by the sigma-clipping algorithm. The pale blue strip drawn underneath the $N$ profile shows the uncertainty associated with each spectral point. The thick green bar on the left side of the spectrum of the $N$ parameter shows the standard deviation of the $N$ profile. Top right panel: linear fit used to determine the magnetic field value using Stokes $V$ (i.e., \bz). The red solid line shows the best fit. From the linear fit we obtain \bz\,=\,431$\pm$57\,G. Bottom right panel: same as the bottom left panel, but for the null profile (i.e., \nz). From the linear fit we obtain \nz\,=\,33$\pm$44\,G.} 
\label{fig:hd46328_bonn} 
\end{center} 
\end{figure*}
\begin{figure*}[ht!]
\begin{center}
\includegraphics[width=0.45\textwidth, angle=0]{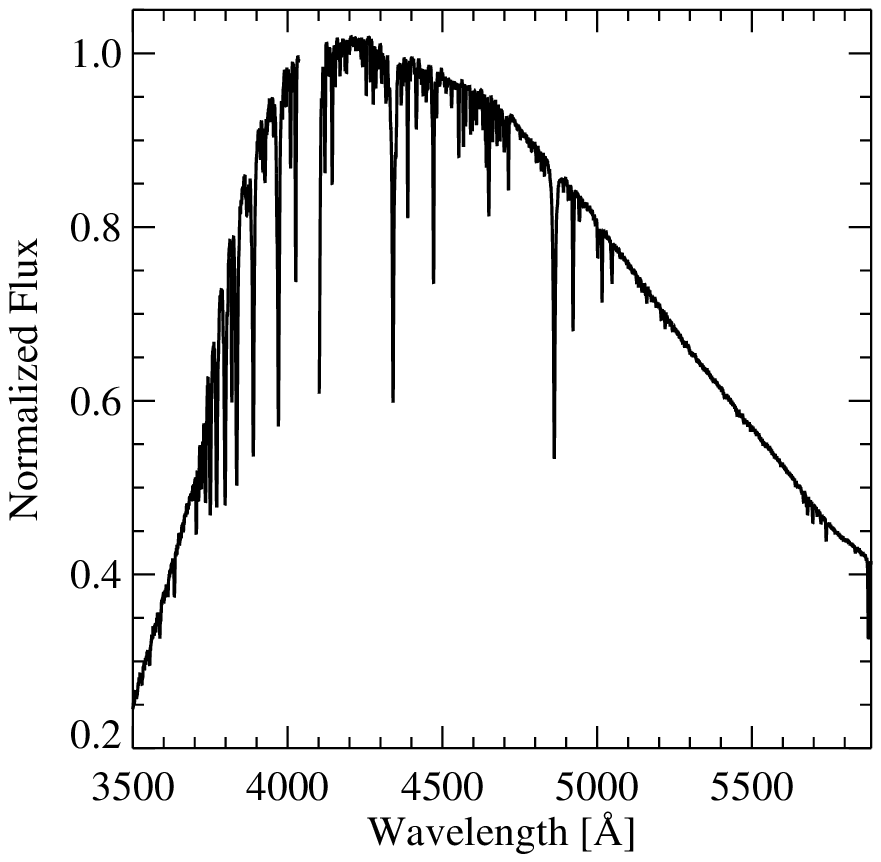}
\includegraphics[width=0.45\textwidth, angle=0]{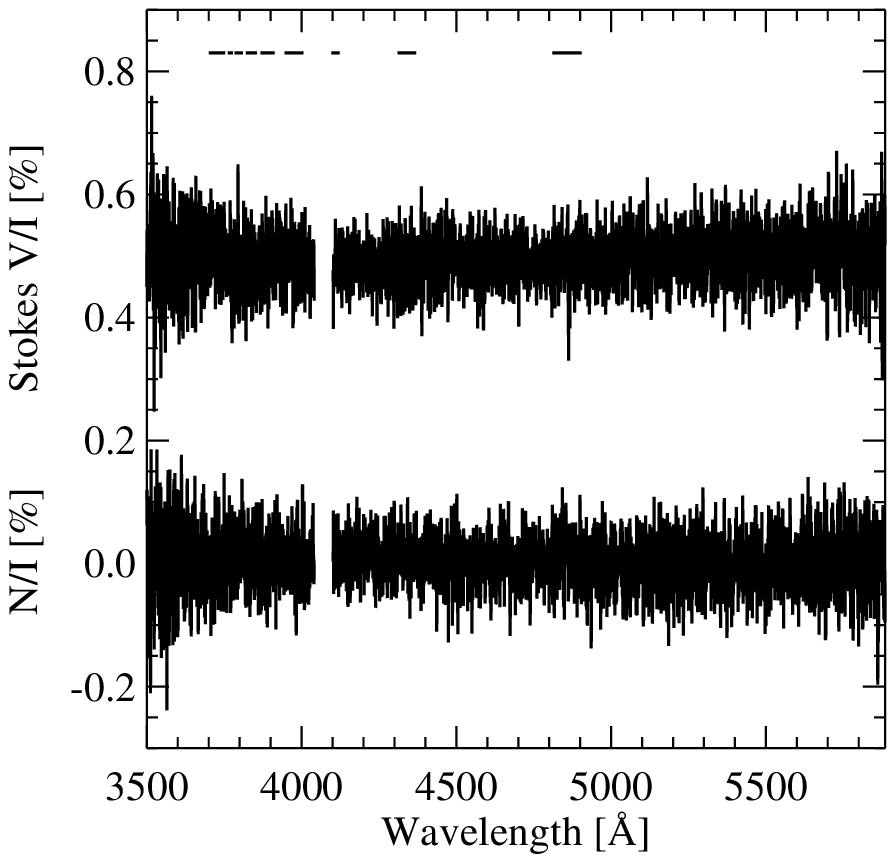}
\includegraphics[width=0.32\textwidth, angle=0]{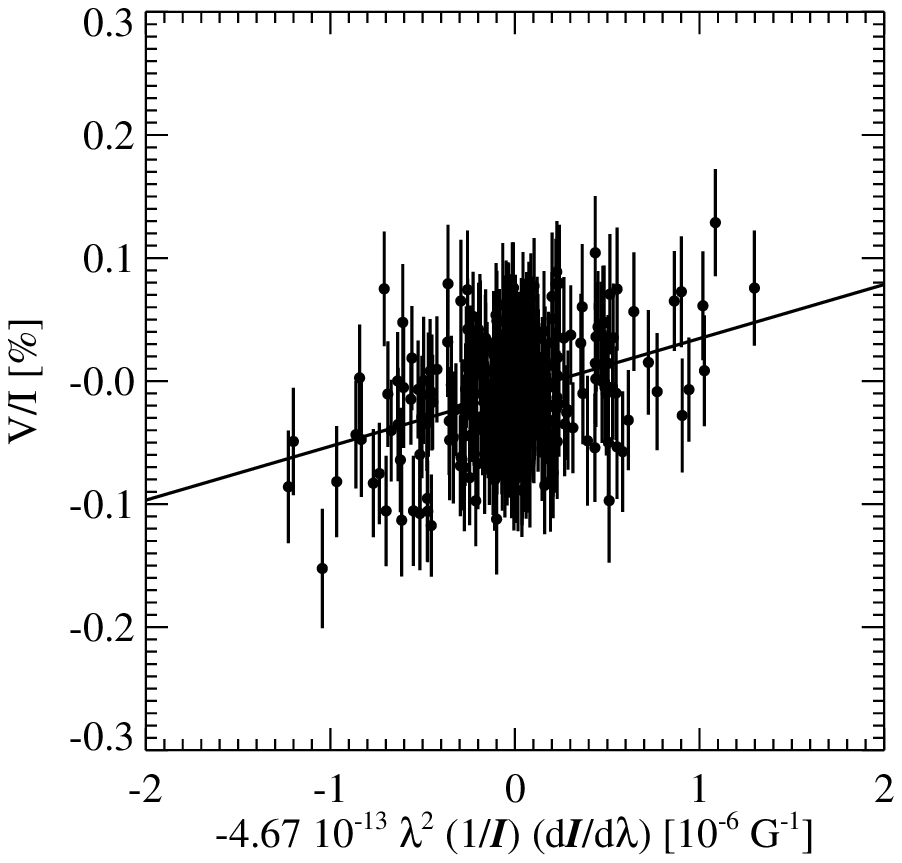}
\includegraphics[width=0.32\textwidth, angle=0]{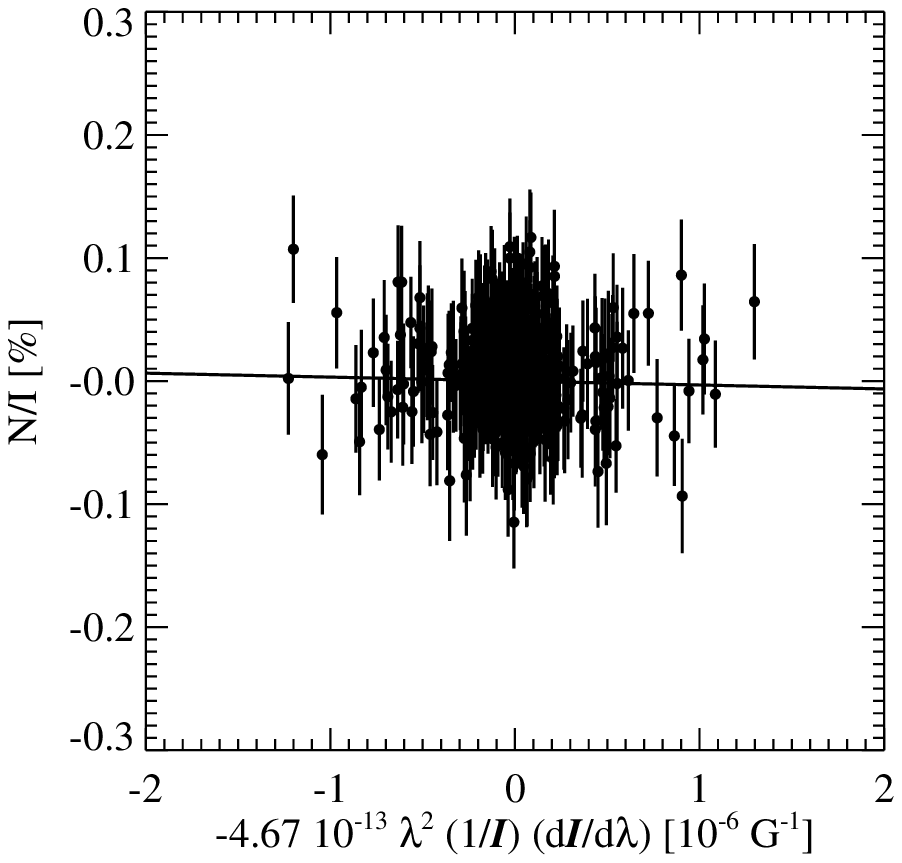}
\includegraphics[width=0.32\textwidth, angle=0]{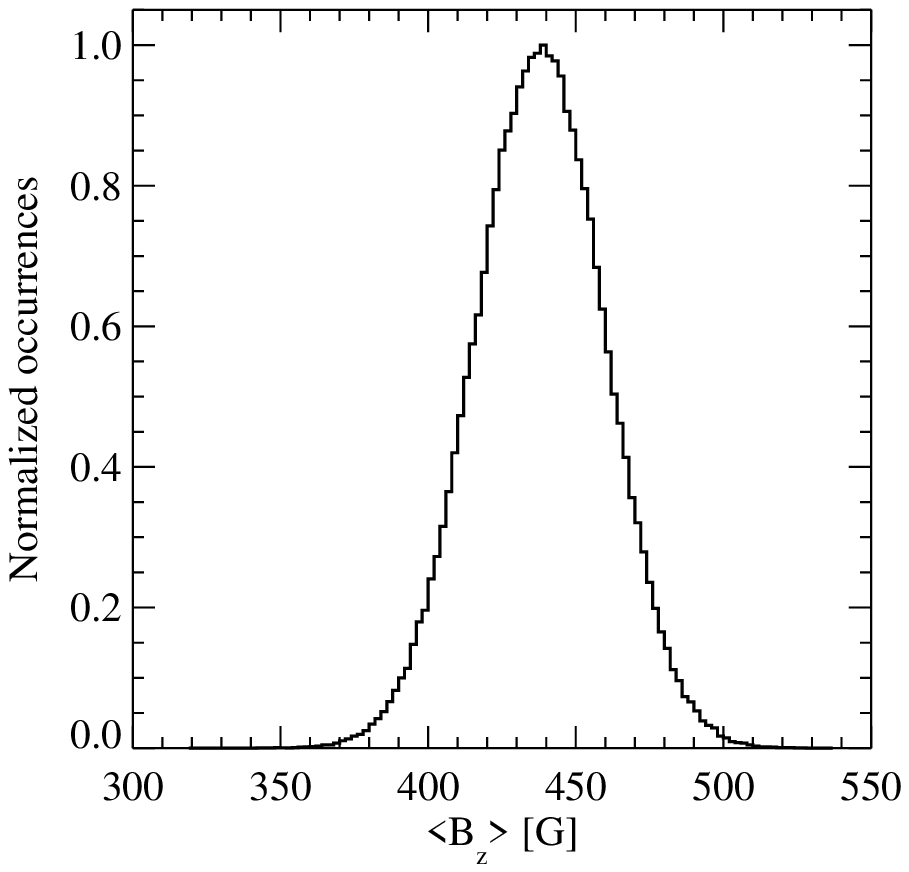}
\caption{Overview of the results of the analysis of the FORS2 data of HD\,46328, collected on 9 April 2013, considering the hydrogen lines, using the Potsdam pipeline. Top left panel: Stokes~$I$ arbitrarily normalised to the highest value. Top right panel: the top profile shows Stokes~$V$ (in \%), while the bottom panel shows the spectrum of the $N$ parameter (in \%). The Stokes~$V$ spectrum is shifted by 0.5 upwards for better visibility. The regions used to calculate the magnetic field are marked by horizontal lines close to the top of the panel. Bottom left panel: linear fit to Stokes~$V$. Bottom middle panel: linear fit to the $N$ spectrum. From the linear fit, we determine $\left<N_{\rm z}\right> = -32\pm61$\,G. Bottom right panel: distribution of the magnetic field values $P(\left<B_{\rm z}\right>)$, obtained via bootstrapping. From the distribution $P(\left<B_{\rm z}\right>)$, we obtain the most likely value for the longitudinal magnetic field $\left<B_{\rm z}\right> = 438\pm60$\,G. We note that the gaps in the region around H$\delta$ in the two upper panels result from masking an internal reflection in that spectral range.}
\label{fig:hd46328_potsdam} 
\end{center} 
\end{figure*}
Within the Potsdam pipeline, the parallel and perpendicular beams were extracted from the raw FORS2 data using a pipeline written in the MIDAS environment by T.~Szeifert. This pipeline reduction by default includes background subtraction and no flat-fielding. A unique wavelength calibration frame was used for each night. The spectra were resampled with a spectral bin size of 0.1\,\AA/pix.

Stokes~$V$ and $I$ were combined in the same way as for the Bonn pipeline. The $V/I$ spectra were rectified using a linear function in the way described by \citet{hubrig2014b}. The diagnostic null spectra, $N$, were calculated as pairwise differences from all available $V$ spectra. From these, 3$\sigma$-outliers were identified and used to clip the $V$ spectra. Following these steps, a visual inspection of all resulting spectra is necessary to ensure that no spurious signals have gone undetected.

Given the Stokes~$I$ and $V$ spectra, the mean longitudinal magnetic field \bz\ is derived for the wavelength region 3645--5880\,\AA\ by linear regression. In the past, the Potsdam pipeline followed the same path as the Bonn pipeline, using Eq.~\ref{eq:chi} and applying the $\chi^2$-correction to the resulting error, if the $\chi^2$ was larger than 1. Since we used 0.1\,\AA/pix as spectral bin size, we had to multiply the resulting error by a factor $\sqrt{7.5}$. Now, we relied on the bootstrapping technique, first introduced by \citet{rivinius2010} for the magnetic field measurements. For this, we generated $M = 250\,000$ statistical variations of the original dataset and analysed the resulting distribution $P(\left<B_{\rm z}\right>)$ of the M regression results, where Eq.~\ref{eq:chi} was applied to each of the statistical variations. Mean and standard deviation of this distribution were identified with the most likely mean longitudinal magnetic field and its 1$\sigma$ error, respectively. The main advantage of this method is that it provides an independent error estimate.
\subsection{Comparison}
Table~\ref{tab:pipelines} summarises the main nominal similarities and differences between the two pipelines. Although both pipelines applied a sigma-clipping algorithm and a normalisation of the Stokes $V$ spectrum and of the $N$ profile, these operations were performed in significantly different ways. The Bonn pipeline used a polynomial to rectify the final co-added Stokes $V$ spectrum and applied the same function to the $N$ profile, while the Potsdam pipeline used a linear function to rectify each single Stokes $V$ spectrum obtained from each pair of frames (i.e., $-$45$^{\circ}$, $+$45$^{\circ}$), with the $N$ profile being the difference of already rectified Stokes $V$ spectra. The Potsdam pipeline applied a sigma clipping algorithm based on deviations from the $N$ profile, similarly to the Bonn pipeline, but because of the oversampling, it also rejected the ten points next to the deviating ones. We considered that for the brightest stars there might be an additional difference in the number of frames considered for the analysis, because of the differences in identifying and discarding saturated frames within the two pipelines, with the Bonn pipeline having a more severe criterion (i.e., a frame is removed when 20 or more neighbouring pixels have a number of counts larger than 60\,000, each). Another substantial difference is in the wavelength ranges selected for the analysis of the spectra using hydrogen lines (or metallic lines) that were manually selected on a star-by-star basis by the users of each pipeline.
\section{Results}\label{results}
\subsection{Magnetic field detection rate}\label{Bfield}
%
\onlfig{
\begin{figure*}[ht!]
\begin{center}
\includegraphics[width=150mm]{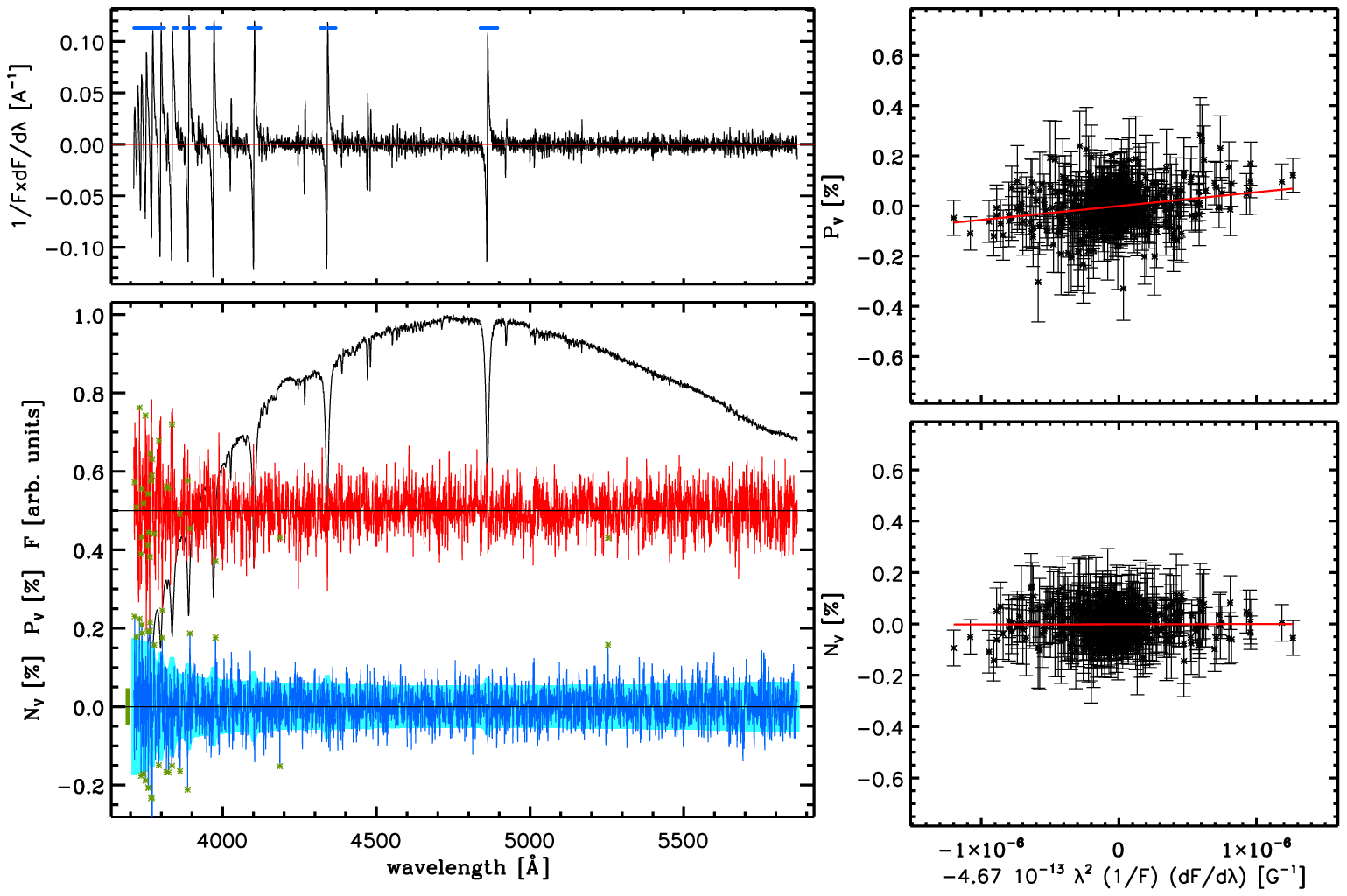}
\caption{Same as Fig.~\ref{fig:hd46328_bonn}, but for the magnetic standard star HD\,125823 observed on the 8th of February 2014. From the linear fit we obtain \bz\,=\,570$\pm$99\,G and \nz\,=\,12$\pm$82\,G.} 
\label{fig:hd125823} 
\end{center} 
\end{figure*}
}
Table~\ref{tab:mag.field} lists the magnetic field values obtained using the two pipelines. Following \citet{bagnulo2012}, the BOB collaboration decided to consider a magnetic field to be detected only above the 5$\sigma$ level and with a \nz\ value consistent with zero. The average S/N of the spectra is about 2500 with an average uncertainty of about 80\,G (considering the measurements conducted on the hydrogen lines), in agreement with the empirical S/N-uncertainty relation given by \citet{bagnulo2015}.

The whole sample is composed of 50 stars (28 O-type stars, 19 B-type stars, 1 A-type supergiant, and 2 F-type stars; note that the spectra of the two stars classified in Simbad as F-type suggest instead an earlier spectral type), two of them being the magnetic reference stars HD\,46328 and HD\,125823. The sample comprises at least three spectroscopic binaries (HD\,164492C, HD\,117357, and HD\,92206c; no high-resolution spectra are available for most of the observed stars, hence only limited information on possible binarity is available), five likely post-main-sequence stars (HD\,168607, HD\,168625, HD\,92207, HD\,72754, and HD\,48279A\,B), and one known chemically peculiar He-rich star (CPD\,$-$57\,3509). Ten stars have a \vsini\ value above $\sim$100\,\kms.

On the basis of this sample, and excluding the two magnetic reference stars, we detected three magnetic stars: HD\,54879, HD\,164492C, and CPD\,$-$57\,3509. The corresponding detection rate is therefore of 6$\pm$4\%, consistent with that obtained by the Magnetism in Massive Stars (MiMeS) survey \citep{wade2014}. By only considering the slow rotators instead, we derive a slightly higher magnetic field detection rate of 8$\pm$5\%, still consistent with that given by the MiMeS survey. Thus, the detection rate amongst slow rotators is apparently only slightly enhanced. This is surprising, given that the bimodal \vsini\ distribution of massive stars \citep[e.g.,][]{dufton2013,oscar2013,iacob} may suggest that about 25\% of the O- and B-type stars show a \vsini\ below 100\,\kms, but about 80\% of the 64 magnetic O- and B-type stars discussed by \citet{petit2013} have a projected rotational velocity below this threshold. Both numbers together lead to an expected detection rate of about 20\% amongst the slow rotators.

The reason for this discrepancy remains unclear at present, but biases could lead to this situation; several magnetic stars have been selected from secondary magnetic field indicators (spectral variability, X-ray emission, etc.), for instance, before their field has been determined, which could imply that the non-biased detection rate is lower than the reported one. Moreover, unlike the intermediate-mass stars, the massive stars appear not to show a magnetic desert \citep{fossati2015}, meaning that many of them could have relatively weak fields that remained undetected. To resolve this puzzle is left to future investigations.

For three stars, HD\,102475, HD\,118198, and HD\,144470, we obtained a measurement of the magnetic field at the 3--4$\sigma$ level using both pipelines, but either from hydrogen lines or the entire spectrum, but never both. Although further FORS2 observations led to clear non-detections, it would be important to observe these stars with a high-resolution spectropolarimeter to perform a deeper search for a magnetic field.
\subsection{Standard stars: HD\,46328 and HD\,125823}
Figures~\ref{fig:hd46328_bonn} and \ref{fig:hd46328_potsdam} illustrate the results obtained for the analysis of the hydrogen lines of the magnetic standard star HD\,46328 from the Bonn and Potsdam pipelines, respectively. Figure~\ref{fig:hd125823} illustrates the results of the Bonn pipeline for the analysis of the hydrogen lines of the magnetic standard star HD\,125823.

The star HD\,46328 ($\xi^1$\,CMa) is a $\beta$\,Cep star \citep{saesen2006} for which the presence of a magnetic field has first been reported by \citet{hubrig2006} and \citet{hubrig2009}. This was further confirmed by high-resolution spectropolarimetry \citep{silvester2009,four2011,shultz2012}. \citet{hubrig2011} used the FORS1 measurements to model the magnetic field of HD\,46328, assuming a dipolar configuration of the magnetic field. They obtained a rotation period of P\,=\,2.17937$\pm$0.00012\,days, a dipolar magnetic field strength B$_{\mathrm d}$ of 5.3$\pm$1.1\,kG, and an obliquity $\beta$ of 79.1$^\circ$$\pm$2.8$^\circ$. As shown in Table~\ref{tab:mag.field}, both pipelines led to the measurement of a positive longitudinal magnetic field (at the $\sim$7$\sigma$ level) of about 400\,G, as expected on the basis of the previous FORS1 measurements. 

Taking advantage of the longer time-base, we used the FORS1 and FORS2 measurements of \bz, obtained from the analysis of the whole spectrum, to improve the estimate of the stellar rotation period. To be consistent with the FORS1 measurements, we used the FORS2 results of the Potsdam pipeline for this analysis. We derived the stellar rotation period adopting the frequency analysis and mode identification for asteroseismology (FAMIAS) package \citep{zima2008} and the phase dispersion minimization (PDM) method \citep{j1971,s1978}, consistently obtaining a period of P\,=\,2.17950$\pm$0.00009\,days. Following \citet{breger1993} we find this period to be significant. On the basis of Musicos and ESPaDOnS high-resolution spectropolarimetric observations, \citet{shultz2015} suggested a rotation period longer than 40\,years. Their measurement of the period is mostly constrained by Musicos observations made at very high airmass, which led to negative values of \bz. We can only report here that the FORS observations conducted in the past years always led to positive values of \bz, and that only further observations obtained in the next 2--5 years will allow unambiguously distinguishing between the two solutions.

\begin{figure}[ht!]
\begin{center}
\includegraphics[width=90mm]{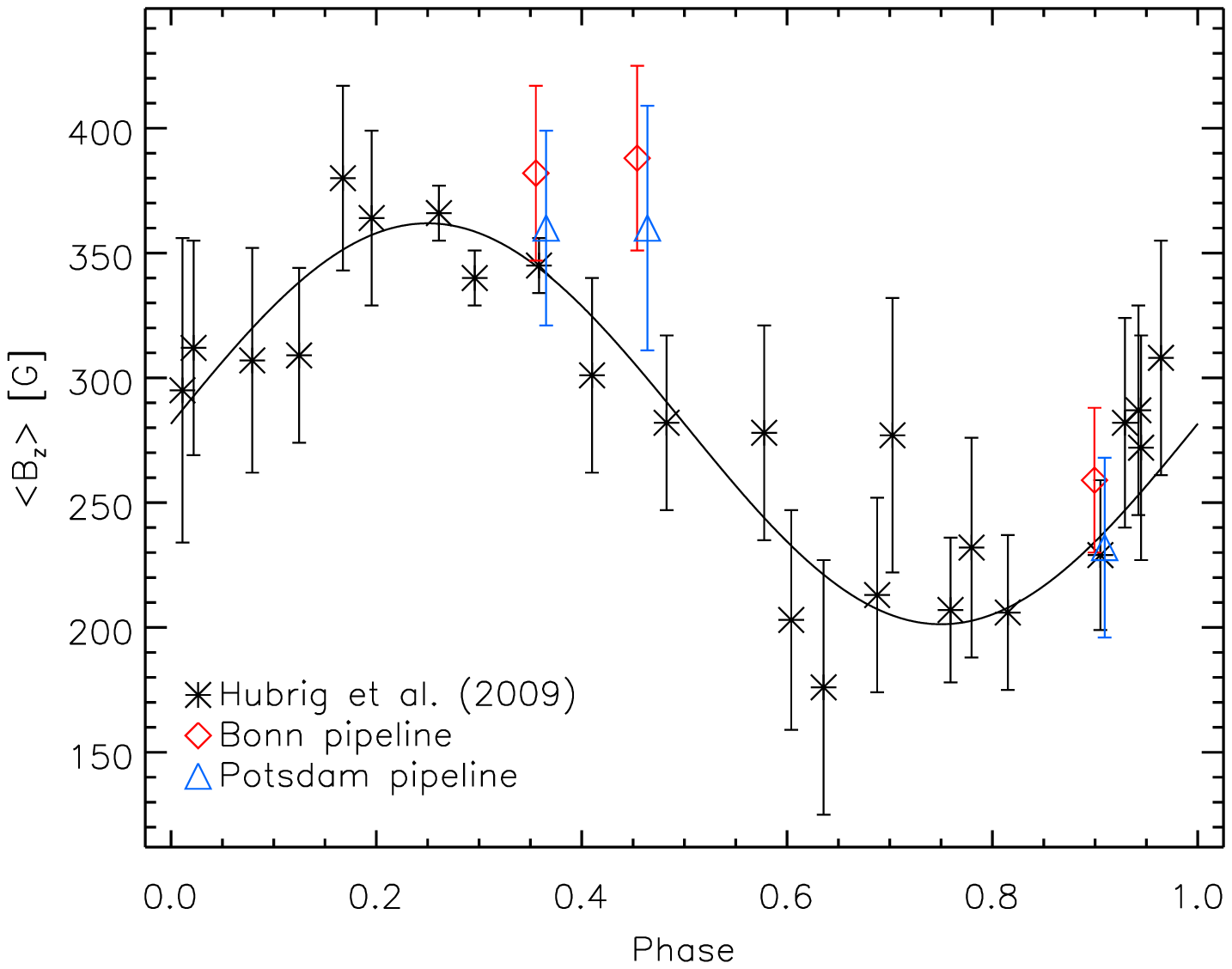}
\caption{Phase plot of the \bz\ values obtained for HD\,46328 from the FORS1 \citep[black asterisks;][]{hubrig2009} and FORS2 (red rhombs: Bonn pipeline, blue triangles: Potsdam pipeline; using the whole spectrum) data, and the sine wave function calculated using the magnetic field model given by \citet{hubrig2011}. A slight phase shift has been applied between our two sets of FORS2 measurements for visualisation purposes.} 
\label{fig:phase_plot_hd46328} 
\end{center} 
\end{figure}

Figure~\ref{fig:phase_plot_hd46328} shows the phase plot obtained using the FORS1 and FORS2 measurements, and the results of the magnetic field modelling given by \citet{hubrig2009}. The results obtained with both pipelines fit the expected behaviour of the longitudinal magnetic field well. This is most likely because the two sets of measurements were obtained with essentially the same instrument (the polarimetric optics of FORS1 were moved to FORS2 after the FORS1 decommissioning) and using similar (almost identical in the case of the Potsdam pipeline) analysis techniques.

\begin{figure}[ht!]
\begin{center}
\includegraphics[width=90mm]{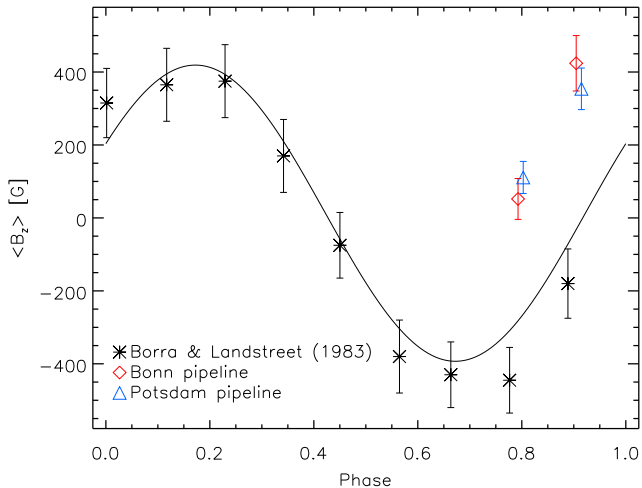}
\caption{Phase plot of the \bz\ values obtained for HD\,125823 from the measurements of \citet{borra1983} (black asterisks) and FORS2 (red rhombs: Bonn pipeline, blue triangles: Potsdam pipeline; using the whole spectrum) data, and the sine wave function calculated using the magnetic field model given by \citet{bychkov2005}. A slight phase shift has been applied between the two sets of FORS2 measurements for visualisation purposes.} 
\label{fig:phase_plot_hd125823} 
\end{center} 
\end{figure}

The star HD\,125823 (a\,Cen) is a Bp star with a rotation period of 8.817744$\pm$0.000019\,days \citep{catalano1996}. \citet{borra1983} detected a magnetic field ranging between $-$470\,G and $+$430\,G. We used the stellar magnetic field model by \citet{bychkov2005} to compare the FORS2 measurements (from both pipelines) with that of \citet{borra1983}. We note that \citet{bychkov2005} considered a period of 8.8171\,days, which is slightly different from that given by \citet{catalano1996}. The phase plot is shown in Fig.~\ref{fig:phase_plot_hd125823}. The FORS2 measurements do not fit the magnetic field model well that was obtained by \citet{bychkov2005} using the results of \citet{borra1983}. This could be due to a systematic shift (of $\sim$400\,G) between the two datasets due to the use of different instruments, setups, and wavelength regions for the magnetic field measurements \citep{landstreet2014}, and/or more likely to small errors in the magnetic model that, given the long time-span between the two sets of observations, led to a significant discrepancy (e.g., a phase shift of $\sim$0.3).
%
\subsection{New detections: HD\,54879, HD\,164492C, and CPD\,$-$57\,3509}
The star HD\,54879 is a single, slowly rotating O9.7V star \citep{sota2011} and a probable member of the CMa\,OB1 association \citep{claria1974}. The discovery of the magnetic field was presented by \citet{castro2015}. Figure~\ref{fig:hd54879} shows the outcome of the Bonn pipeline indicating the clear detection of the magnetic field at the $\sim$9$\sigma$ level, already reported by \citet{castro2015}. The stellar photospheric spectrum does not present any morphological peculiarity, typical for example of Of?p stars, and its analysis did not reveal any chemical peculiarity. The only distinctive feature in the spectrum of HD\,54879 is a prominent H$\alpha$ emission that \citet{castro2015} attributed to circumstellar material, as the comparison of the H$\alpha$ line profile with that of the star defining the O9.7V spectral type excludes the stellar wind as the cause of the emission.
\onlfig{
\begin{figure*}[ht!]
\begin{center}
\includegraphics[width=150mm]{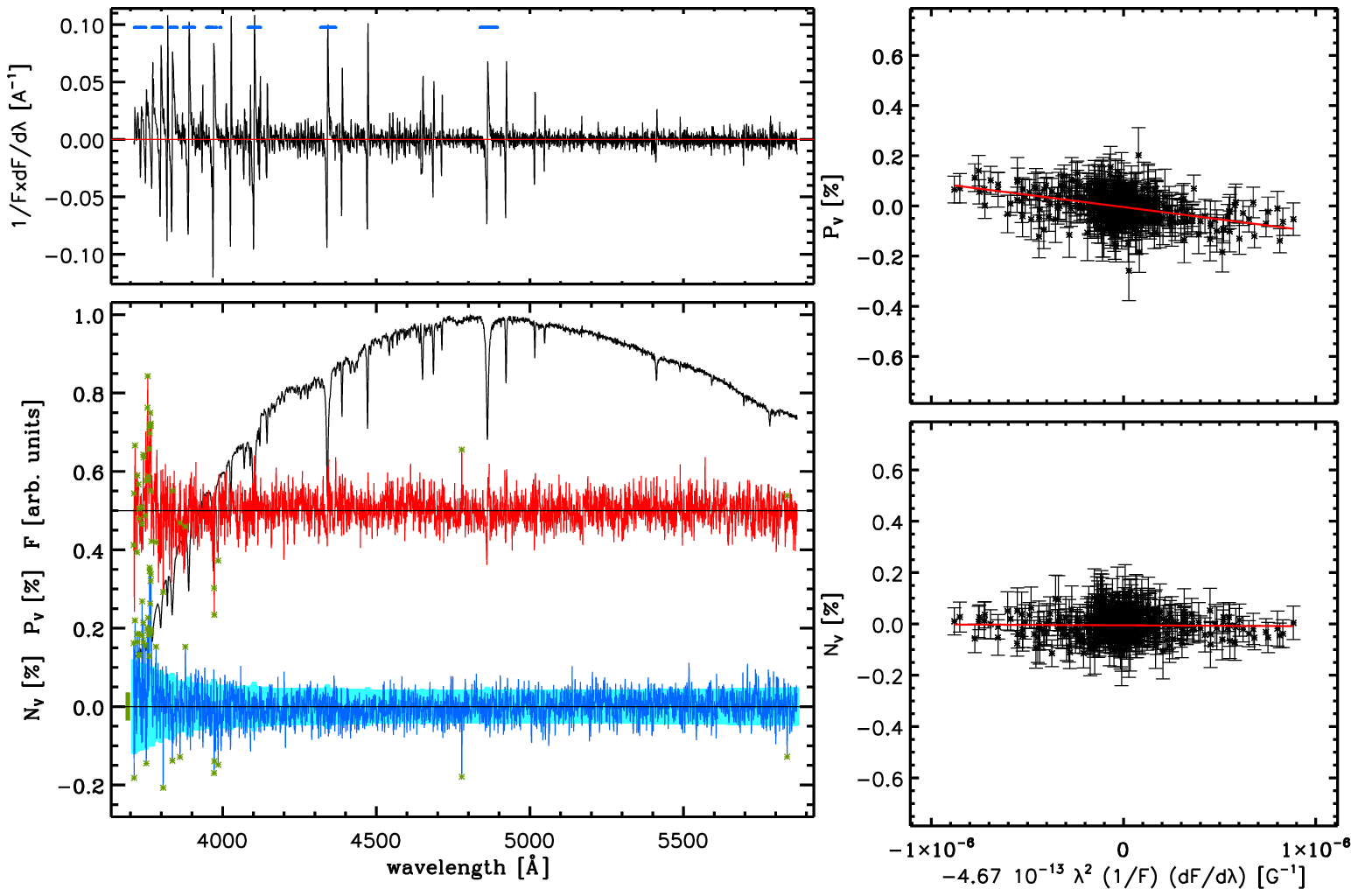}
\caption{Same as Fig.~\ref{fig:hd46328_bonn}, but for HD\,54879 observed on the 8th of February 2014. From the linear fit we obtain \bz\,=\,$-$978$\pm$88\,G and \nz\,=\,$-$36$\pm$76\,G.} 
\label{fig:hd54879} 
\end{center} 
\end{figure*}
}

The star HD\,164492C is a massive star in the centre of the Trifid nebula. \citet{hubrig2014} reported the detection of a rather strong magnetic field on the basis of FORS2 and HARPSpol data. Figure~\ref{fig:hd164492C} illustrates the clear detection of the magnetic field at the $\sim$9$\sigma$ level, already reported by \citet{hubrig2014}\footnote{Note that there is a slight difference between the \bz\ and \nz\ measurements reported here (Table~\ref{tab:mag.field}) and that given by \citet{hubrig2014} because of a more recent update in the Bonn pipeline.}. The high-resolution HARPSpol observations and further high-resolution UVES spectra revealed that HD\,164492C is in fact a multiple system, composed of at least two stars. More details about this system and the UVES observations will be given in a follow-up paper (Gonz{\'a}lez et al., in prep.).
\onlfig{
\begin{figure*}[ht!]
\begin{center}
\includegraphics[width=150mm]{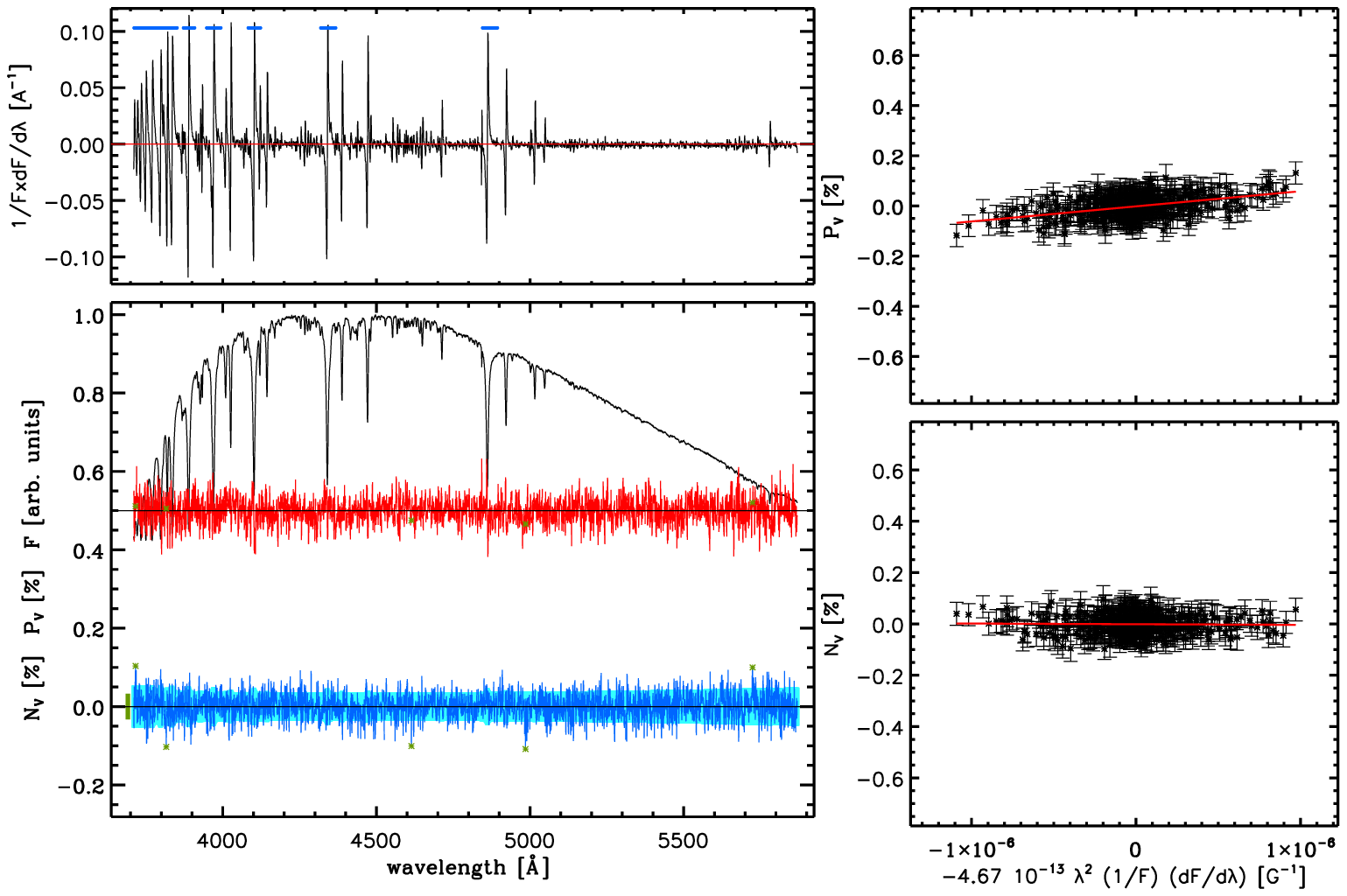}
\caption{Same as Fig.~\ref{fig:hd46328_bonn}, but for HD\,164492C observed on the 8th of April 2013. From the linear fit we obtain \bz\,=\,602$\pm$54\,G and \nz\,=\,$-$25$\pm$53\,G.} 
\label{fig:hd164492C} 
\end{center} 
\end{figure*}
}

The star CPD\,$-$57\,3509 is a He-rich B2 star member of the $\sim$10\,Myr old open cluster NGC\,3293. We observed the star with FORS2 twice during the run in February 2014. Figure~\ref{fig:cpd-573509} reveals the detection of the magnetic field (at the $\sim$5$\sigma$ level) obtained from the data collected on 7 February 2014. Following the FORS2 measurements, we observed the star with the HARPSpol high-resolution spectropolarimeter confirming the presence of a magnetic field. Our measurements of the magnetic field are suggestive of the presence of a rather strong and rapidly varying magnetic field. A preliminary \nlte\ analysis confirms the He-rich nature of the star (about three times solar). Its membership in the NGC\,3293 open cluster allows us to conclude that the star has evolved throughout about one third of its main-sequence lifetime. This makes CPD\,$-$57\,3509 one of the most evolved He-rich stars with a tight age constraint, promising to provide information on the evolution of stars with magnetically confined stellar winds. More details will be given in a dedicated paper (Przybilla et al., in prep.).
\onlfig{
\begin{figure*}[ht!]
\begin{center}
\includegraphics[width=150mm]{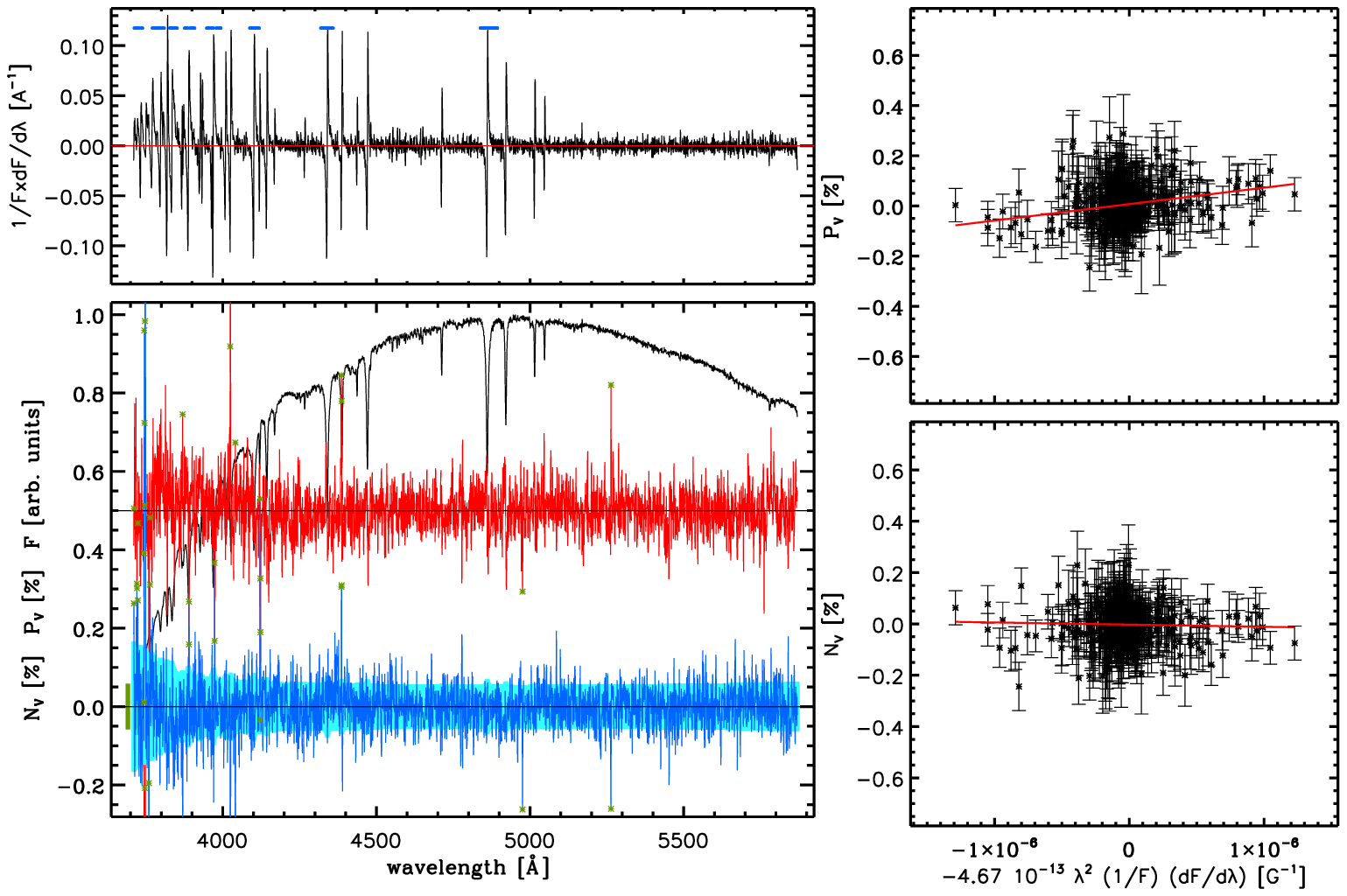}
\caption{Same as Fig.~\ref{fig:hd46328_bonn}, but for CPD\,$-$57\,3509 observed on the 7th of February 2014. From the linear fit we obtain \bz\,=\,659$\pm$109\,G and \nz\,=\,$-$120$\pm$97\,G.} 
\label{fig:cpd-573509} 
\end{center} 
\end{figure*}
}
%
\section{Discussion}\label{sec:discussion}
%
\begin{figure*}[ht!]
\begin{center}
\includegraphics[width=185mm]{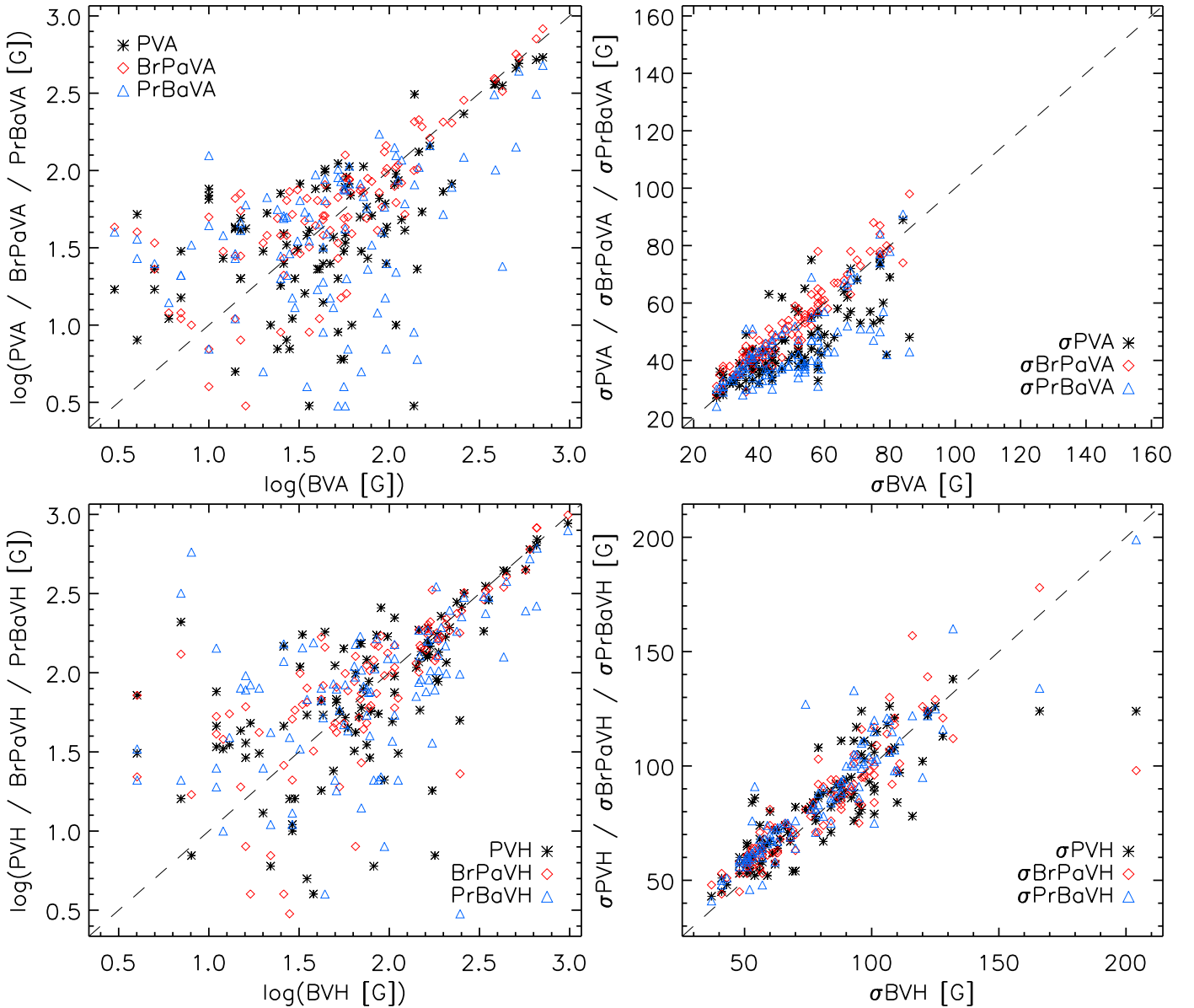}
\caption{Top left panel: comparison between the |\bz| values obtained by analysing the whole spectrum with the Bonn pipeline (BVA) and i) the Potsdam pipeline (PVA, black asterisks), ii) the Potsdam pipeline, but reducing the data with the Bonn pipeline (BrPaVA, blue triangles), iii) the Bonn pipeline, but reducing the data with the Potsdam pipeline (PrBaVA, red rhombs). Top right panel: same as top left panel, but for the uncertainties on the \bz\ values. Bottom left panel: same as top left panel, but for the |\bz| values obtained analysing the hydrogen lines. Bottom right panel: same as bottom left panel, but for the uncertainties on the \bz\ values. The meaning of each acronym used in the labels and legends of each panel (e.g., BVH) can be found in the header of Tables~\ref{tab:mag.field} and \ref{tab:cross.check}.}
\label{fig:plottoneV} 
\end{center} 
\end{figure*}
\begin{figure*}[ht!]
\begin{center}
\includegraphics[width=185mm]{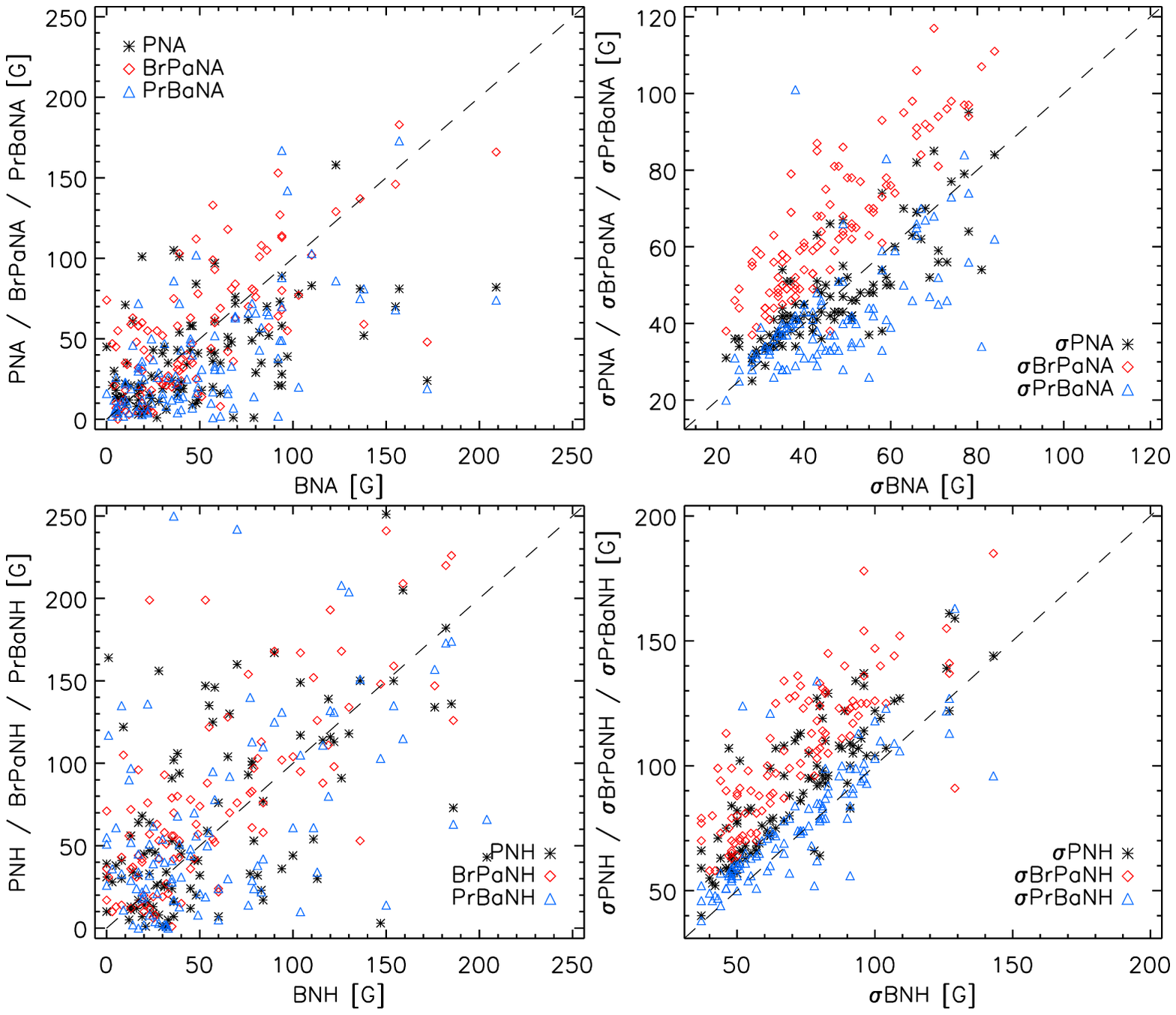}
\caption{Same as Fig.~\ref{fig:plottoneV}, but for the \nz\ values.} 
\label{fig:plottoneN} 
\end{center} 
\end{figure*}

One of the characteristics of the BOB collaboration is that the reduction and analysis of the spectropolarimetric data is independently carried out by two teams using different and independent tools and pipelines. This gives us the possibility to directly compare the results on a statistically large sample of stars. 

To make a more thorough comparison, we also applied a mixed reduction and analysis of the data: we derived the \bz\ and \nz\ values using the Bonn pipeline for the data reduction (i.e., bias subtraction, spectral extraction, wavelength calibration) and the Potsdam pipeline for the spectral analysis (i.e., derivation of the Stokes parameters and of the magnetic field values), and vice versa. The results of this test are presented in Table~\ref{tab:cross.check}.

Figures~\ref{fig:plottoneV} and \ref{fig:plottoneN} show the comparison between the results obtained by reducing and analysing the spectra (hydrogen lines or whole spectrum) with the Bonn and Potsdam pipelines, or the mixed reduction and analysis. We consider here 102 sets of measurements, each set composed of four measurements (i.e., \bz\ and \nz\ obtained from the analysis of the hydrogen lines or of the whole spectrum), and obtained in four different ways with six possible comparisons (i.e., BrPa, PrBa, and PrPa compared to BrBa; BrPa and PrBa compared to PrPa; BrPa compared to PrBa -- the meaning of each acronym can be found in the header of Tables~\ref{tab:mag.field} and \ref{tab:cross.check}), for a total of 2448 direct comparisons.

Figures~\ref{fig:plottoneV} and \ref{fig:plottoneN} display a general good agreement among the four sets of results, and for most cases ($\sim$96.73\%) the differences are within 2$\sigma$. In about 1.6\% of the cases the difference between the various sets of \bz\ and \nz\ values is above 3$\sigma$. This is close to the expectations of Gaussian statistics. In addition, \citet{bagnulo2012} showed that even slight changes in just one step in the data reduction or analysis procedure may lead to variations in the \bz\ and \nz\ values of 2--3$\sigma$. We note that the comparison of the uncertainties shown in Fig.~\ref{fig:plottoneV} and \ref{fig:plottoneN} is slightly affected by the fact that the Potsdam pipeline calculates the uncertainties using the nominal CCD gain, while the uncertainties calculated with the Bonn pipeline, because of the $\chi^2$ scaling, account for deviations from the nominal value of the CCD gain.

The best agreement is found when comparing the results of the two pipelines separately (i.e., BrBa vs. PrPa) and of each pipeline with what is obtained from the mixed Bonn pipeline reduction and Potsdam pipeline analysis (i.e., BrBa vs. BrPa and PrPa vs. BrPa) with $<$2\% of the cases having a difference larger than 2$\sigma$. For the other three comparisons (i.e., BrBa vs. PrBa, PrPa vs. PrBa, and PrBa vs. BrPa), in 5--8\% of the cases the difference is larger than 2$\sigma$, about what expected by random noise. These results do not seem to display a regular pattern that would allow one to conclude anything about the relative importance of the adopted reduction or analysis procedure in the final results. 

The largest differences ($\geq$4$\sigma$) instead follow a clear pattern as they are found almost exclusively among the measurements conducted for the magnetic stars. This is probably because, for the non-magnetic stars, both \bz\ and \nz\ measure noise, for which one may expect a Gaussian behaviour, which therefore leaves limited room for large deviations. On the other hand, for the magnetic stars, uncertainties are generally small and differences in the data reduction or analysis procedure may indeed modify the Stokes $V$ signatures, which therefore leads to significant differences. This suggests that the optimal data reduction and analysis procedure may therefore be sought by considering magnetic (standard) stars \citep[see also][]{landstreet2014} in addition to the analysis of large samples \citep[see e.g.,][]{bagnulo2012,bagnulo2015}. The identification of the exact reduction step(s) leading to the observed differences is beyond the scope of this work.

On the basis of our analysis, we conclude that except for a few cases \citep[e.g., HD\,92207;][]{bagnulo2013}, the several discrepancies reported in the literature are mostly due to the interpretation of the significance of the results, that is, whether 3--4$\sigma$ detections are considered as genuine or not.
\section{Conclusion}\label{sec:conclusion}
Within the context of the BOB collaboration, whose primary aim is characterising the incidence of magnetic fields in slowly rotating massive stars, we obtained FORS2 spectropolarimetric observations of a set of 50 massive stars selected considering their spectral type, luminosity class, and projected rotational velocity. Within this sample, we also observed two massive stars that were previously known to host a magnetic field and that we used as standards (HD\,46328 and HD\,125823). The observations were performed in April 2013 and February 2014. 

We derived the longitudinal magnetic field values using two fully independent reduction and analysis pipelines to compare the results and decrease the probability of spurious detections. We detected the magnetic field for both HD\,46328 and HD\,125823. We used previous FORS1 measurements, in addition to our FORS2 results, to further constrain the rotation period of HD\,46328, obtaining a best fit of P\,=\,2.17950$\pm$0.00009\,days. We did not find evidence for a long rotation period ($>$40\,years), as recently suggested by \citet{shultz2015}, but only further observations obtained in the next years will allow unambiguously distinguishing between the two solutions. Our FORS2 results are also a good fit to the magnetic field model of HD\,46328 presented by \citet{hubrig2011}. In contrast, our measurements do not fit the magnetic field model of HD\,125823 well that was reported by \citet{bychkov2005} on the basis of measurements obtained by \citet{borra1983}, possibly because of systematic shifts between the two datasets \citep[see e.g.,][]{landstreet2014} and/or of small errors in the magnetic field model that would be magnified when considering measurements so much spread in time.

Within the remaining sample of 50 stars, we detected a magnetic field for three of them: HD\,54879, HD\,164492C, and CPD\,$-$57\,3509. For the chemically normal O9.7V star HD\,54879 we detected a longitudinal magnetic field with a maximum strength of about 1\,kG \citep[see][for more details]{castro2015}. HD\,164492C is a massive binary system in the centre of the Trifid nebula for which we detected a magnetic field of about 600\,G, although it is unclear which of the stars composing this system is magnetic \citep[see][for more details]{hubrig2014}. The star CPD\,$-$57\,3509 is a He-rich B2 star member of the NGC\,3293 open cluster. We detected a rapidly varying longitudinal magnetic field of about 700\,G, further confirmed by follow-up HARPSpol high-resolution spectropolarimetric observations (Przybilla et al., in prep.).

Considering the whole sample of observed stars, but excluding HD\,46328 and HD\,125823, we obtained a magnetic field detection rate of 6$\pm$4\%, while by considering only the apparently slow rotators we reached a slightly higher detection rate of 8$\pm$5\%. Both numbers are comparable to the magnetic field incidence rate of O- and B-type stars of 7\% reported by \citet{wade2014}. Given that the vast majority of magnetic massive stars rotate slowly, we expected to find a higher magnetic fraction (about 20\%) from our sample of slow rotators. That this is not so may hint at biases in the magnetic stars sample and might imply that a large number of massive stars contain magnetic fields that are too weak to be detected at present \citep{fossati2015}.

Finally, we compared the magnetic field values obtained from the two reduction and analysis pipelines. We obtained a general good agreement, and for only about 1\% of the cases, the difference is above 3$\sigma$, the majority of those being for the magnetic stars. Our results indicate that most discrepancies on magnetic field detections reported in the literature are mainly caused by the interpretation of the significance of the results, that is, it depends on whether 3--4$\sigma$ detections are considered as genuine, or not.
\begin{acknowledgements}
LF acknowledges financial support from the Alexander von Humboldt Foundation. TM acknowledges financial support from Belspo for contract PRODEX GAIA-DPAC. LF thanks Stefano Bagnulo and Konstanze Zwintz for fruitful discussions. SH and MS thank Thomas Szeifert for providing the pipeline for the FORS spectra extraction. We thank the referee, Gautier Mathys, for his useful comments. This research has made use of the SIMBAD and ViZieR databases, and of the WEBDA database, operated at the Department of Theoretical Physics and Astrophysics of the Masaryk University.
\end{acknowledgements}
%
%




\end{document}